\documentclass[aip, jmp, amsmath, amssymb, preprint]{revtex4-1}

\usepackage{epsfig}

\usepackage{graphicx}
\usepackage{dcolumn}
\usepackage{bm}

\usepackage{braket} 
\usepackage{graphicx}

\begin{document}

\title{Asymptotic Analysis of the Wigner $3j$-Symbol in the Bargmann Representation}

\author{Liang Yu}

\affiliation{ Department of Physics, University of California, Berkeley, California 94720 USA}

\date{\today} 

\begin{abstract}
We derive the leading asymptotic limit of the Wigner $3j$-symbol from a stationary phase approximation of a twelve dimensional integral, obtained from an inner product between two exact Bargmann wavefunctions. We show that, by the construction of the Bargmann inner product, the stationary phase conditions have a geometric description in terms of the Hopf fibration of ${\mathbb C}^6$ into ${\mathbb R}^3 \times {\mathbb R}^3 \times {\mathbb R}^3$. In addition, we find that, except for the usual modification of the quantum numbers by $1/2$, the imaginary part of the logarithm of a Bargmann wavefunction, evaluated at the stationary points, is equal to the asymptotic phase of the $3j$-symbol. 
\end{abstract}
\pacs{03.65.Sq, 02.30.Fn, 02.20.-a} 

\maketitle

\section{Introduction}

This paper presents a new derivation of the asymptotic limit of the Wigner $3j$-symbol, using stationary phase approximation on an inner product of wavefunctions in the Bargmann representation\cite{bargmann1961, bargmann1962, bargmann1967}. The asymptotic limit of the $3j$-symbol was first derived in 1968 by Ponzano and Regge\cite{ponzano-regge-1968}. Since then, there were several subsequent derivations \cite{neville1971, miller1974, schulten1975a, schulten1975b, reinsch1999, littlejohn2007}. Despite the long history behind this result, this paper uncovers some new connections between the asymptotic formula and the Bargmann representation\cite{bargmann1961, bargmann1962, bargmann1967}, which is an exact representation of $SU(2)$ by holomorphic functions.

We find that, by construction of the inner product in the Bargmann representation, the stationary phase conditions are formulated in terms of the projections of three vectors in a triangle through the Hopf fibration. In addition, the imaginary part of the logarithms of the Bargmann wavefunctions is equal to the asymptotic phase of the $3j$-symbol, albeit, without the usual modification of the quantum numbers by $1/2$. Finally, we point out that the asymptotic wavefunctions appearing in the recent study of the Wigner $3nj$-symbols\cite{roberts1999, charles2008, littlejohn2007, littlejohn2010a, yu2011a, yu2011b, yu2011c} are actually Bargmann wavefunctions in disguise.

We now give an outline of the paper. In section \ref{ch2: sec_bargmann_wavefunction}, we give some background on the Bargmann representation, and write down the exact Bargmann wavefunctions for the eigenstates in the definition of the $3j$-symbol. In section \ref{ch2: sec_integral_rep}, we express the $3j$-symbol as a twelve-dimensional integral. We then apply a stationary phase approximation to this integral to derive the asymptotic limit of the $3j$-symbol in section \ref{ch2: sec_stationary}. After the derivation of the asymptotic formula, we relate the Bargmann wavefunctions to the holomorphic wavefunctions used in geometric quantization in section \ref{ch2: sec_rel_geometricQ}, and to and the multidimensional WKB wavefunctions in section \ref{ch2: sec_rel_schwinger}. The last section contains some comments and conclusions.

\section{\label{ch2: sec_bargmann_wavefunction}The Bargmann Wavefunctions}

Bargmann constructed a unitary representation of the quantum rotation group $SU(2)$ in a series of papers \cite{bargmann1961, bargmann1962, bargmann1967} in the 1960s. Bargmann's construction used Schwinger's construction for the generators of $SU(2)$ from two sets of commuting boson operators. Instead of using the quantum harmonic oscillators, however, Bargmann implemented the boson operators in terms of multiplication and differentiation in a complex coordinate. The result was a Hilbert space that consisted entirely of holomorphic functions.

We now briefly describe the construction of the Bargmann representation. Let $z_\mu$, $\mu = 1,2$, denote coordinates  on ${\mathbb C}^2$. The bosonic commutation relations,

\begin{equation}
[ z_\mu , \frac{\partial }{\partial z_\mu} ] = 1  \, , 
\end{equation}
allow us to form two independent sets of boson creation operators $z_\mu$ and annihilation operators $\partial_{z_\mu}$, $\mu=1,2$. Each set of boson operators has a number operator,

\begin{equation}
\hat{N}_\mu = z_\mu \partial_{z_\mu} \, ,
\end{equation}
whose eigenvalues are integers, and whose eigenfunctions are monomials in $z_{\mu}$. Following Schwinger's construction of the generators of $SU(2)$ in terms of boson creation and annihilation operators, we define the $SU(2)$ generators in the Bargmann representation by 

\begin{equation}
\label{ch2: eq_J_i_operators}
\hat{J}_i = z_\mu \sigma^i_{\mu\nu} \frac{\partial}{\partial z_\nu} \, , 
\end{equation}
where $\sigma^i$ are the Pauli matrices. These generators satisfy the usual $SU(2)$ Lie algebra commutation relations, 

\begin{equation}
[ \hat{J}_i , \hat{J}_j ] = i \epsilon_{ijk} \hat{J}_k \, . 
\end{equation}
The Casimir operator is 

\begin{equation}
\label{ch2: eq_J_square}
\hat{J}^2 = \hat{I} ( \hat{I} + 1 ) \, , 
\end{equation}
where $\hat{I}$ is given by

\begin{equation}
\label{ch2: eq_number_operator}
\hat{I} = \frac{1}{2} (\hat{N}_1 + \hat{N}_2) \, .
\end{equation}
Since the eigenvalues of $\hat{N}_1, \hat{N}_2$ are integers, the eigenvalues of $\hat{I}$ are half-integers $j=(n_1 + n_2)/2$, and the eigenvalues of $\hat{J}^2$ are given by $j (j+1)$.

The Hilbert space for a single Bargmann space, ${\mathcal F}_2$, consists of holomorphic functions on ${\mathbb C}^2$. That is, 

\begin{equation}
{\mathcal F}_2 = \{ f \in {\mathcal C}^{\infty} ( {\mathbb C}^2 ) \, | \, \braket{f | f} < \infty  \} \, ,
\end{equation}
where the inner product is given by 

\begin{equation}
\label{ch2: eq_inner_prod}
\braket{f | g} = \int_{{\mathbb C}^2} \overline{f}(z) \, g(z) \, \frac{e^{-|z|^2}}{\pi^2} \, d^2 z  \,  ,
\end{equation}
and where $d^n z$ denotes the usual Euclidean measure on ${\mathbb C}^2 = {\mathbb R}^{4}$.

In the recoupling theory of three angular momenta, the Hilbert space is the tensor product of three copies of the Hilbert space for a single Bargmann space. The wavefunctions are holomorphic functions on ${\mathbb C}^6 = {\mathbb C}^2 \times {\mathbb C}^2 \times {\mathbb C}^2$. Let us denote the coordinates by $(z_{r\mu}) = (z_{11}, z_{12}, z_{21}, z_{22}, z_{31}, z_{32})$, $r = 1,2,3$, $\mu=1,2$. Let us denote the operators in Eq.\ (\ref{ch2: eq_J_i_operators}) and (\ref{ch2: eq_number_operator}) that act on the $r$th angular momentum space by $\hat{I}_r$ and $\hat{J}_{ri}$, respectively.

We now find the eigenstates that appear in the definition of the $3j$-symbol. The basis state $\psi_{jm}(z) = \braket{z| j_1 m_1 j_2 m_2 j_3 m_3}$ satisfies the following eigenvalue equations and normalization condition, 

\begin{eqnarray}
\label{ch2: eq_eig_eqn1}
\hat{I}_r \psi_{jm} (z) &=& \frac{1}{2} (\hat{N}_{r1} + \hat{N}_{r2})  \psi_{jm} (z)  =  j_r \psi_{jm} (z) \, ,  \\  
\label{ch2: eq_eig_eqn2}
\hat{J}_{rz} \psi_{jm} (z) &=& \frac{1}{2} (\hat{N}_{r1} - \hat{N}_{r2})  \psi_{jm} (z)  = m_r \psi_{jm} (z) \, ,  \\
\braket{\psi_{jm} | \psi_{jm} } &=& 1\, , 
\end{eqnarray}
for $r = 1,2,3$. The eigenvalue equations determine the degrees of the monomial in $z_{r\mu}$, and the normalization condition fixes the constant in front. We find

\begin{equation}
\label{ch2: eq_jm_state}
\psi_{jm}(z_1, z_2) = \prod_{r = 1}^3  \frac{1}{\sqrt{(j_r-m_r)! (j_r+m_r)!}} \, z_{r1}^{j_r+m_r} \, z_{r2}^{j_r-m_r} \, . 
\end{equation}

We now find the rotationally invariant wavefunction $\psi_{\rm inv}(z) = \braket{z| j_1 j_2 j_3 {\mathbf 0}}$, which satisfies the following eigenvalue equations and normalization condition, 

\begin{eqnarray}
\label{ch2: eq_eig_eqn3}
\hat{I}_r \psi_{\rm inv} (z) &=&  \frac{1}{2} (\hat{N}_{r1} + \hat{N}_{r2})  \psi_{\rm inv} (z) = j_r \psi_{\rm inv} (z) \, ,  \\
\label{ch2: eq_eig_eqn4}
\hat{J}_i \psi_{\rm inv} (z)  &\equiv& \left( \sum_r \hat{J}_{ri} \right) \psi_{\rm inv} (z) = 0 \, ,  \\
\braket{\psi_{\rm inv}  | \psi_{\rm inv}  } &=& 1\, ,
\end{eqnarray}
for $r = 1,2,3$, and $i = 1,2,3$. 

We use the diagonal $SU(2)$ action generated by the generators $\hat{J}_{i}$ to find $\psi_{\rm inv} (z)$. Let us denote the diagonal group action by $T_U$, where $U \in SU(2)$. Its action on a Bargmann wavefunction $f(z)$ is given by 

\begin{equation}
T_U (f(z)) = f(U^\dagger_{\mu\nu}z_{1\nu}, U^\dagger_{\mu\nu}z_{2\nu}, U^\dagger_{\mu\nu}z_{3\nu}) \, . 
\end{equation}
Since $U$ is unitary, the three determinants 

\begin{equation}
\delta_1 = z_{21}z_{32} - z_{31} z_{22}  \, , \quad 
\delta_2 =  z_{31}z_{12} - z_{11} z_{32} \, , \quad 
\delta_3 =  z_{11}z_{22} - z_{21} z_{12}  \, , 
\end{equation}
are invariant under $T_U$. Using these determinants to construct $\psi_{\rm inv}(z)$, we find 

\begin{equation}
\label{ch2: eq_wigner_state}
\psi_{\rm inv} (z) = \frac{\delta_1^{k_1} \delta_2^{k_2} \delta_3^{k_3}}{\sqrt{(j_1 + j_2 + j_3)! k_1! k_2! k_3 !}} \, , 
\end{equation}
where the non-negative integers $k_1, k_2, k_3$, which are chosen to satisfy the eigenvalue equation in Eq. (\ref{ch2: eq_eig_eqn3}). They are given by

\begin{equation}
\label{ch2: eq_k_values}
k_1 = j_2 + j_3 - j_1 \, ,  \quad 
k_2 = j_3 + j_1 - j_2 \, ,  \quad 
k_3 = j_1 + j_2 - j_3 \, .
\end{equation}

\section{\label{ch2: sec_integral_rep}An Integral Representation of the $3j$-Symbol}

Taking the scalar product between $\psi_{jm}(z)$ from Eq.\ (\ref{ch2: eq_jm_state}) and $\psi_{\rm inv} (z)$ from Eq.\ (\ref{ch2: eq_wigner_state}), we obtain an exact integral representation of the $3j$-symbol,

\begin{eqnarray}
\label{ch2: eq_3j_integral}
 \left( 
\begin{array}{ccc}
j_1 & j_2 & j_3 \\
m_1 & m_2 & m_3 \\
\end{array}
\right)    
&=& \braket{j_1 m_1 j_2 m_2 j_3 m_3 | j_1 j_2 j_3 {\mathbf 0} }    \\  \nonumber
&=&  N \int_{{\mathbb C}^6} \, d^{12}\, z \, 
\overline{z}_{11}^{j_1+m_1} \,
\overline{z}_{12}^{j_1-m_1} \,
\overline{z}_{21}^{j_2+m_2} \,
\overline{z}_{22}^{j_2-m_2} \,
\overline{z}_{31}^{j_3+m_3} \,
\overline{z}_{32}^{j_3-m_3} \,
 \, \, \,  e^{-|z|^2} \,    \nonumber   \\   \nonumber
&& (z_{21}z_{32} - z_{31}z_{22})^{k_1} \, 
(z_{31}z_{12} - z_{32}z_{11})^{k_2} \, 
(z_{11}z_{22} - z_{12}z_{21})^{k_3} \,  , 
\end{eqnarray}
where the constant in front is

\begin{eqnarray}
N &=& \frac{1}{\pi^6 [ (j_1+m_1)!  (j_1-m_1)!  (j_2+m_2)! 
(j_2-m_2)! (j_3+m_3)! (j_3-m_3)! ]^{1/2} }    \nonumber   \\
&& \quad \frac{1}{[ (j_1 + j_2 + j_3)! \, k_1! \, k_2! \, k_3! ]^{1/2} } \, , 
\label{ch2: eq_constant_2}
\end{eqnarray}
and where $k_1, k_2, k_3$ are given in Eq.\ (\ref{ch2: eq_k_values}).

\section{\label{ch2: sec_stationary}Stationary Phase Approximation}

We now apply the stationary phase approximation to the integral expression in Eq.\ (\ref{ch2: eq_3j_integral}) in the limit that $j_i$ for $i = 1,2,3$ are large. The basic formula for stationary phase approximation is given by

\begin{equation}
\label{ch2: eq_stationary_formula}
\int e^{f(x)} d^n x = (2 \pi)^{n/2} \sum_p \frac{e^{ f(p) } }{\sqrt{- {\rm Hess}_p (f) }}  \,  ,
\end{equation}
where $p$ stands for the stationary phase points that satisfy the twelve stationary phase conditions 

\begin{equation}
\label{ch2: eq_stationary_condition}
\partial_{z_{r\mu}} f(z, \overline{z}) = 0 \, , \quad \partial_{\overline{z}_{r\mu}} f(z, \overline{z}) = 0 \, . 
\end{equation}
In the denominator of Eq.\ (\ref{ch2: eq_stationary_formula}), ${\rm Hess}_p(f)$ is the determinant of second derivatives of $f$ evaluated at $p$. In the general case that the stationary phase points $p$ are not isolated points, the sum in Eq.\ (\ref{ch2: eq_stationary_formula}) is replaced by an integral over the set of stationary phase points.

\subsection{\label{ch2: subsec_stationary_points}Stationary Phase Points}

In the case of the $3j$-symbol in Eq.\ (\ref{ch2: eq_3j_integral}), the phase function $f$ in Eq.\ (\ref{ch2: eq_stationary_formula})  is 

\begin{equation}
\label{ch2: eq_phase_function}
f(z, \overline{z}) = \ln \overline{\psi}_{jm} (z) + \ln \psi_{\rm inv}(z)  - \sum_{r \mu} | z_{r\mu} |^2  \, .
\end{equation}
To find the stationary phase points, we use the fact that the Bargmann representation of the generators of $SU(2)$ are also differential operators. That is, the twelve conditions (\ref{ch2: eq_stationary_condition}) on the first derivatives of $f$ imply 

\begin{eqnarray}
\hat{I}_r^* \,  f(z, \overline{z}) = 0  \, ,  \quad
\hat{J}_{rz}^* \, f(z, \overline{z}) = 0   \, ,  \quad 
\hat{I}_r  \, f(z, \overline{z}) = 0  \, ,  \quad  
\hat{J}_i  \,  f(z, \overline{z}) = 0  \, .
\end{eqnarray}
Here we pick the Bargmann operators that appear in the eigenvalue equations (\ref{ch2: eq_eig_eqn1}), (\ref{ch2: eq_eig_eqn2}), (\ref{ch2: eq_eig_eqn3}), (\ref{ch2: eq_eig_eqn4}). This way, we can use the eigenvalue equations to easily find the result of the Bargmann operators on the first two terms of $f$. We rewrite the eigenvalue equations, 

\begin{eqnarray}
\hat{I}_r^* \overline{\psi}_{jm}  = j_r \overline{\psi}_{jm}    \, , \quad
\hat{J}_{rz}^* \overline{\psi}_{jm}  = m_r \overline{\psi}_{jm}     \, ,  \quad
\hat{I}_r \psi_{\rm inv}  = j_r \psi_{\rm inv}   \, , \quad
\hat{J}_i \psi_{\rm inv} = 0     \, ,  
\end{eqnarray}
into the following form, 

\begin{equation}
\label{ch2: eq_bargmann_op_result1}
\hat{I}_r^* \ln \overline{\psi}_{jm} (\overline{z})  = j_r \, , \quad  
\hat{J}_{rz}^* \ln \overline{\psi}_{jm}  (\overline{z}) = m_r \, , \quad  
\hat{I}_r \, \ln \psi_{\rm inv} (z) = j_r \, , \quad 
\hat{J}_i  \ln \psi_{\rm inv}  (z) = 0 \, .
\end{equation}
Since ${\overline{\psi}}_{jm}(\overline{z})$ is anti-holomorphic, $\hat{I}_r$ and $\hat{J}_i$ annihilate $ \ln {\overline{\psi}}_{jm}(\overline{z})$. Similarly, since $\psi_{\rm inv}(z)$ is holomorphic, $\hat{I}_r^*$ and $\hat{J}_{rz}^*$ annihilate  $ \ln \psi_{\rm inv}(z)$.

Finally, we apply these Bargmann operators to the last remaining term $\sum_{r \mu} |z_{r\mu}|^2$ in $f$. The result is

\begin{eqnarray}
I_r  & \equiv &  \hat{I}_r^* \sum_{r \mu} |z_{r\mu}|^2 =   \frac{1}{2} \,
 \sum_{\mu}
z_{r\mu}  \,  
\overline{z}_{r \mu}    \, ,  \quad \quad 
J_{rz}   \equiv   \hat{J}_{rz}^* \sum_{r \mu} |z_{r\mu}|^2 = \frac{1}{2} \sum_{\mu\nu}
\overline{z}_{r\mu}  \,  (\sigma_z)_{\mu\nu}
z_{r\nu}     \, ,  \\
I_r  & \equiv & \hat{I}_r \sum_{r \mu} |z_{r\mu}|^2 =  \frac{1}{2}  \,
 \sum_{\mu}
z_{r\mu}  \,  
\overline{z}_{r \mu}    \, ,   \quad \quad
J_i \,  \equiv  \hat{J}_i  \sum_{r \mu} |z_{r\mu}|^2 = \frac{1}{2} \,
 \sum_{r \mu\nu}
z_{r\mu}  \,  (\sigma^i)_{\mu\nu}
\overline{z}_{r \nu}    \, .
\label{ch2: eq_bargmann_op_result2}
\end{eqnarray}
The functions $I_r$, $J_{rz}$, and $J_i$ are functions of the Hopf map $J_{ri}: {\mathbb C}^2 \times {\mathbb C}^2 \times {\mathbb C}^2  \rightarrow {\mathbb R}^3 \times {\mathbb R}^3 \times {\mathbb R}^3 $, given by

\begin{eqnarray}
\label{ch2: eq_Hopf_Jrx}
J_{rx} &=& \frac{1}{2} \sum_{\mu\nu} \overline{z}_{r\mu}  \,  (\sigma_x)_{\mu\nu} z_{r\nu}  
= \frac{1}{2} \, (\overline{z}_{r1} z_{r2} + \overline{z}_{r2} z_{r1}) = {\rm Re} \, (\overline{z}_{r1} z_{r2}) \, ,  \\
J_{ry} &=& \frac{1}{2} \sum_{\mu\nu} \overline{z}_{r\mu}  \,  (\sigma_y)_{\mu\nu} z_{r\nu}  
= \frac{1}{2} \, (\overline{z}_{r1} z_{r2} - \overline{z}_{r2} z_{r1}) = {\rm Im} \, (\overline{z}_{r1} z_{r2}) \, ,  \\
J_{rz} & =& \frac{1}{2} \sum_{\mu\nu} \overline{z}_{r\mu}  \,  (\sigma_z)_{\mu\nu} z_{r\nu}  
= \frac{1}{2} \, (|z_{r1}|^2 - |z_{r2}|^2) \, .
\label{ch2: eq_Hopf_Jrz}
\end{eqnarray}

Putting together the result of apply the Bargmann operators to $f$, Eqs.\ (\ref{ch2: eq_bargmann_op_result1}) - (\ref{ch2: eq_bargmann_op_result2}), the stationary phase conditions become

\begin{equation}
\label{ch2: eq_stationary_phase_condition}
I_r = j_r  \, , \quad \quad J_{rz} = m_r  \, , \quad \quad  J_i = 0 \, .
\end{equation}

Using the Hopf map, we can interpret the above conditions in ${\mathbb R}^3 \times {\mathbb R}^3 \times {\mathbb R}^3$, which we will call the angular momentum space. These conditions are geometrical conditions on three vectors ${\bf J}_1, {\bf J}_2, {\bf J}_3$. The conditions $I_r = j_r$ state the vectors ${\bf J}_r$  have lengths $j_r$, $r = 1,2,3$, respectively. The conditions $J_i = 0$ state that the vectors ${\bf J}_1, {\bf J}_2, {\bf J}_3$, put together head to tail, form a triangle. The conditions $J_{rz} = m_r$ fix the $z$ projections of the three vectors.

Thus a natural way to find the solutions to Eq.\ (\ref{ch2: eq_stationary_phase_condition}) in ${\mathbb C}^6$ is to first find three vectors ${\bf J}_r$ in ${\mathbb R}^3$ that satisfy the geometric conditions described above. After the vectors are found, we lift each vector up to a spinor $z_{r\mu}$ in the Hopf fiber above. This procedure is carried out in Appendix \ref{ch2: sec_stationary_spinors}, and is very similar to the calculations in the multidimensional WKB approach\cite{littlejohn2007}, except the length of the vectors in that paper are $j_r+1/2$ instead of $j_r$. The stationary phase point $p$ from Appendix \ref{ch2: sec_stationary_spinors} is given by 

\begin{eqnarray}
\label{ch2: eq_final_spinor_1}
\left( 
\begin{array}{c}
z_{11}  \\
z_{12} 
\end{array}  \right)
&=& \sqrt{2j_1}
\left(
\begin{array}{c}
e^{-i \gamma /2} \cos \beta /2 \cos \eta_2 /2 - e^{i \gamma/2} \sin \beta /2 \sin \eta_2 /2  \\
e^{-i \gamma /2} \sin \beta /2 \cos \eta_2 /2 + e^{i \gamma/2} \cos \beta /2 \sin \eta_2 /2
\end{array}  \right) \, ,   \\  
\label{ch2: eq_final_spinor_2}
\left( 
\begin{array}{c}
z_{21}  \\
z_{22} 
\end{array}  \right)
&=& \sqrt{2j_2}
\left(
\begin{array}{c}
e^{-i \gamma /2} \cos \beta /2 \cos \eta_1 /2 + e^{i \gamma/2} \sin \beta /2 \sin \eta_1 /2 \\
e^{-i \gamma /2} \sin \beta /2 \cos \eta_1 /2 - e^{i \gamma/2} \cos \beta /2 \sin \eta_1 /2 
\end{array}  \right) \, ,   \\  
\label{ch2: eq_final_spinor_3}
\left( 
\begin{array}{c}
z_{11}  \\
z_{12} 
\end{array}  \right)
&=& e^{- i \gamma /2 } \sqrt{2j_3}
\left(
\begin{array}{c}
\cos \beta /2  \\
\sin \beta /2 
\end{array}  \right) \, .
\end{eqnarray}
Here the angles $\eta_1, \eta_2, \beta, \gamma$ are defined in the Eqs.\ (\ref{ch2: eq_eta}), (\ref{ch2: eq_beta}), and (\ref{ch2: eq_gamma}).

Once we have the solution $p$ with $\gamma > 0$, we can find another solution $p'$ given by $\gamma = - | \gamma |$.  These two solutions are not isolated in the solution set. To see that, note any set of three vectors related to the projection of $p$ in the angular momentum space by a rotation about the $z$ axis will continue to satisfy Eqs.\ (\ref{ch2: eq_stationary_phase_condition}). In addition, multiplication by an overall phase for each of the three spinors also preserve the conditions in Eqs.\ (\ref{ch2: eq_stationary_phase_condition}). Thus, the stationary phase points consist of two disjoint $4$-tori, generated by an overall rotation about the $z$-axis and three overall phases of the three spinors.

\subsection{\label{ch2: subsec_3j_formula}The Asymptotic Formula}

After integrating over the two $4$-tori of stationary phase points, and applying the stationary phase approximation to the integral for the $3j$-symbol along the remaining eight directions transversal to the stationary phase points, we find

\begin{eqnarray}
\label{ch2: eq_formula_1}
\left( 
\begin{array}{ccc}
j_1 & j_2 & j_3 \\
m_1 & m_2 & m_3 \\
\end{array}
\right)  
\approx
N \, (2\pi)^8 \,  e^{{\rm Re}  \; f(p)} \; \left[ \frac{e^{i \; {\rm Im} \; f(p)}}{\sqrt{- {\rm Hess}_p(f)}} +  \frac{e^{ i \; {\rm Im} \; f(p')}}{\sqrt{ {\rm Hess}_{p'}(f)}} \right]  \, .
\end{eqnarray}
Here ${\rm Hess}_p(f)$ is the determinant of an $8 \times 8$ matrix of second derivatives along the eight transversal directions. In Eq.\ (\ref{ch2: eq_formula_1}), one factor of $(2\pi)^4$ comes from integrating along the four angular directions along the stationary points, another factor of $(2\pi)^4$ comes from doing the stationary phase approximation along the eight transversal directions.

The calculation for the Hessian is straightforward but long, so we leave the details of the calculation to Appendix \ref{ch2: sec_hessian}. Inserting the result, Eq.\ (\ref{ch2: eq_Hessian}), into Eq.\ (\ref{ch2: eq_formula_1}), we find

\begin{eqnarray}
\label{ch2: eq_formula_2}
\left(
\begin{array}{ccc}
j_1 & j_2 & j_3 \\
m_1 & m_2 & m_3
\end{array}
\right)
&=& \frac{(2 \pi)^8 N e^{{\rm Re} f_1(p)}}{2^7}  \frac{e^{i ( {\rm Im} f_1(p) + \frac{\pi}{4}) }+ e^{i ( {\rm Im} f_1(p') - \frac{\pi}{4} )} }{\sqrt{ \Delta_z}} \,  \\  \nonumber
&=& \frac{(2 \pi)^8 N e^{{\rm Re} f_1(p)}}{2^6  \sqrt{ \Delta_z}} \cos \left( S + \frac{\pi}{4} \right),
\end{eqnarray}
The function $f_1$ has the same functional form as the phase function $f$, but with $j_r$ replaced by $j_r + 1/2$. This usual modification of the quantum numbers comes from the Hessian. See Eq.\ (\ref{ch2: eq_Hessian}). Explicitly, 

\begin{eqnarray}
\label{ch2: eq_phase_function_1}
f_1(z, \overline{z}) &=& \sum_{r=1}^3 \, [ (j_r + 1/2 + m_r) \ln {\overline{z}_{r1}} + (j_r + 1/2 - m_r) \ln \overline{z}_{r2} ]    \\   \nonumber
&&  + \sum_{i=1}^3 (k_i + 1/2) \ln \delta_i  - \sum_{r \mu} | z_{r\mu} |^2  \, .
\end{eqnarray}
In the second equality in Eq.\ (\ref{ch2: eq_formula_2}), we have defined $S = {\rm Im} f_1 (p)$, and used the fact that ${\rm Im} f_1(p') = - {\rm Im} f_1(p)$. Here $\Delta_z$ is given in Eq.\ (\ref{ch2: eq_delta_z}), and is the projected area of the triangle onto the $xy$-plane. Thus, the asymptotic phase $S$ of the $3j$-symbol is equal to the imaginary part of the logarithm of the Bargmann wavefunctions, modulo the modification of the quantum numbers $j_r$ by $1/2$. 

\begin{eqnarray}
\label{ch2: eq_f1}
S &=& \sum_{r=1}^3 [ (j_r + 1/2 + m_r) \arg (\overline{z}_{r1}) + (j_r + 1/2 - m_r) \arg (\overline{z}_{r2}) ]  \\   \nonumber
&& +  \sum_{i=1}^3  (k_i + 1/2)  \arg \delta_i   \, . 
\end{eqnarray}
Because $\delta_i$, $i=1,2,3$ are invariant under overall $SU(2)$ rotations, the last three terms can be evaluated at any point related to the stationary phase points by an overall $SU(2)$ rotation. We choose the reference spinor in Eq.\ (\ref{ch2: eq_ref_spinor}). The spinors at this point are all real, so the last three terms vanish. This is possible because the three vectors form a triangle and can be rotated into the $xz$ plane. The remaining terms evaluated at $p$, is given by

\begin{eqnarray}
\label{ch2: eq_3j_relative_phase}
S &=& J_1 \cos^{-1} \left( \frac{J_1 \cos \beta - m_1 \cos \eta_2}{\sin \eta_2 \, J_{1 \perp} }  \right) + J_2 \cos^{-1} \left( \frac{m_2 \cos \eta_1 - J_2 \cos \beta}{\sin \eta_1 J_{2 \perp} } \right)  \nonumber    \\     \nonumber
&& \, + J_3 \cos^{-1} \left( \frac{J_1 \cos \beta \cos \eta_2 - m_1 }{J_1 \sin \beta \sin \eta_2} \right) + m_1 \cos^{-1} \left( \frac{J_1 \cos \eta_2 - m_1 \cos \beta}{\sin \beta J_{1 \perp} } \right) \\
&& \,  - m_2 \cos^{-1} \left( \frac{J_2 \cos \eta_1 - m_2 \cos \beta}{ \sin \beta \, J_{2 \perp} }  \right) \, , 
\end{eqnarray}
where $J_{r \perp} = \sqrt{J_r^2 - m_r^2}$, and $J_r = j_r + 1/2$. We now calculate the constant factor $(2 \pi)^8 N e^{{\rm Re} f(p)} / 2^6$ in Eq.\ (\ref{ch2: eq_formula_2}). Here 

\begin{eqnarray}
e^{{\rm Re} f} &=& \left( \prod_r |z_{r1}|^{j_r + m_r+1/2} |z_{r2}|^{j_r - m_r+1/2} \right)   \, \, \,  e^{- \sum_{r, \mu} |z_{r \mu}|^2} \,    \\   \nonumber
&& |z_{21}z_{32} - z_{31}z_{22}|^{k_1+1/2} \, 
 |z_{31}z_{12} - z_{32}z_{11}|^{k_2+1/2} \, 
|z_{11}z_{22} - z_{12}z_{21}|^{k_3+1/2} \, . 
\end{eqnarray}
We evaluate $e^{{\rm Re} f}$ at the stationary point $p$. From the conditions $|z_{r1}| = j_r + m_r$, $|z_{r2}| = j_r - m_r$, and $m_1 + m_2 + m_3 = 0$, we find 

\begin{eqnarray}
&& \left( \prod_r |z_{r1}|^{j_r + m_r+1/2} |z_{r2}|^{j_r - m_r+1/2} \right)   \, \, \,  e^{- \sum_{r, \mu} |z_{r \mu}|^2}  \nonumber  \\   \label{ch2: eq_constant_1}
&=& e^{-2 (j_1 + j_2 + j_3) } \sqrt{ \prod_{r} |j_r + m_r|^{j_r + m_r + 1/2} \, |j_r - m_r|^{j_r - m_r + 1/2} }  \, . 
\end{eqnarray}
The remaining factor,  

\begin{equation}
\label{ch2: eq_constant_3}
|z_{21}z_{32} - z_{31}z_{22}|^{k_1+1/2} \, 
 |z_{31}z_{12} - z_{32}z_{11}|^{k_2+1/2} \, 
|z_{11}z_{22} - z_{12}z_{21}|^{k_3+1/2} \, ,
\end{equation}
is invariant under the diagonal $SU(2)$ actions, so we can again evaluate it at the spinors in Eq.\ (\ref{ch2: eq_ref_spinor}). The result for the three factors are

\begin{eqnarray}
\quad  |z_{21}z_{32} - z_{31}z_{22}|^{k_1+1/2}
&=& \left| \sqrt{4 j_2 j_3} \, \sin ( \eta_1 / 2 ) \right|^{k_1 + 1/2}    \nonumber  \\   \nonumber 
&=& \sqrt{ \left| 4 j_2 j_3 \, ( 1  -  \cos \eta_1 ) \right|^{k_1 + 1/2} } \quad \quad \quad    \\   \nonumber 
&=& \sqrt{ \left| 2 j_2 j_3 ( 1  -  \frac{j_1^2 - j_2^2 - j_3^2}{2 j_2 j_3} ) \right|^{k_1 + 1/2} }  \\  
&=& \sqrt{ (j_1 + j_2 + j_3)^{k_1 + 1/2} ( j_2 + j_3 - j_1 )^{k_1 + 1/2} }  \, ,
\end{eqnarray}
\begin{eqnarray}
|z_{31}z_{12} - z_{32}z_{11}|^{k_2+1/2} 
&=& \left| \sqrt{4 j_1 j_3} \, \sin ( \eta_2 / 2 ) \right|^{k_2 + 1/2}  \nonumber \\  \nonumber
&=& \sqrt{ \left| 4 j_1 j_3 \, ( 1  -  \cos \eta_2 ) \right|^{k_2 + 1/2} }     \\  \nonumber
&=& \sqrt{ \left| 2 j_1 j_3 ( 1  -  \frac{j_2^2 - j_1^2 - j_3^2}{2 j_1 j_3} ) \right|^{k_2 + 1/2} }  \\
&=& \sqrt{ (j_1 + j_2 + j_3)^{k_2 + 1/2} ( j_1 + j_3 - j_2 )^{k_2 + 1/2} }  \, , 
\end{eqnarray}
\begin{eqnarray}
|z_{11}z_{22} - z_{12}z_{21}|^{k_3+1/2}   
&=& \left| \sqrt{4 j_1 j_2} \, ( \sin ( \eta_1 / 2 )\, \cos( \eta_2 /2 ) + \sin ( \eta_2 / 2) \, \cos ( \eta_1 / 2 ) ) \right|^{k_3 + 1/2}  \nonumber  \\   \nonumber
&=&  \left| \sqrt{ 4 j_1 j_2 } \, \sin ( (\eta_1 + \eta_2) / 2 ) \right|^{k_3 + 1/2}   \\  \nonumber
&=& \sqrt{ \left| 2 j_1 j_2 \, ( 1  -  \cos ( (\eta_1 + \eta_2)  ) ) \right|^{k_3 + 1/2} }   \\   \nonumber
&=& \sqrt{ \left| 2 j_1 j_2 \, ( 1  -  \cos ( \eta_3  ) ) \right|^{k_3 + 1/2} }  \\   \nonumber
&=& \sqrt{ \left| 2 j_1 j_2 ( 1  -  \frac{j_3^2 - j_1^2 - j_2^2}{2 j_1 j_2} ) \right|^{k_3 + 1/2} }  \\
&=& \sqrt{ (j_1 + j_2 + j_3)^{k_3 + 1/2} ( j_1 + j_2 - j_3 )^{k_3 + 1/2} }   \, . 
\end{eqnarray}
Multiplying together the above results for Eq.\ (\ref{ch2: eq_constant_1}) and Eq.\ (\ref{ch2: eq_constant_3}), we find 

\begin{eqnarray}
e^{{\rm Re} f(p)}   \nonumber
&=&  e^{-2\, (j_1 + j_2 + j_3)}  \, \sqrt{ \prod_i (j_i + m_i)^{j_i+m_i+1/2} \, (j_i - m_i)^{j_i - m_i + 1/2} } \\   \nonumber
&&  \times  \sqrt{ (j_1 + j_2 + j_3)^{j_1 + j_2 + j_3 + 3/2} k_1^{k_1+1/2} k_2^{k_2+1/2}  k_3^{k_3+1/2}}   \\   \nonumber
&\approx&  \sqrt{ \frac{ (j_1+m_1)! (j_1-m_1)! (j_2+m_2)!    
(j_2-m_2)! (j_3+m_3)! (j_3-m_3)!  }{(2\pi)^5} }  \\
&& \times  \sqrt{(j_1 + j_2 + j_3)! k_1! k_2! k_3! ( j_1 + j_2 + j_3 )}  \, , 
\end{eqnarray}
where we have used Stirling's approximation for factorials, $n! \approx (2 \pi)^{1/2} n^{n+1/2} e^{-n}$. Most of the factorials cancel those that appear in $N$ in (\ref{ch2: eq_constant_2}). We find

\begin{equation}
\label{ch2: eq_constant_in_front}
\frac{(2 \pi)^8 N e^{{\rm Re} f(p)}}{2^6} \approx \frac{1}{\sqrt{2 \pi}}  \, . 
\end{equation}

Substituting Eq.\ (\ref{ch2: eq_3j_relative_phase}) and (\ref{ch2: eq_constant_in_front}) into Eq.\ (\ref{ch2: eq_formula_2}), we find the asymptotic formula for the Wigner $3j$-symbol is 

\begin{equation}
\left(
\begin{array}{ccc}
j_1 & j_2 & j_3 \\
m_1 & m_2 & m_3
\end{array}
\right)
= \pm \frac{\cos (S + \pi/4)}{\sqrt{ 2 \pi | \Delta_z | }} \, .
\end{equation}

\section{\label{ch2: sec_rel_geometricQ}Bargmann Wavefunctions and Geometric Quantization}

Recently, Roberts\cite{roberts1999} and Charles \cite{charles2008} provided two independent derivations of the Ponzano-Regge formula for the Wigner $6j$-symbol using wavefunctions from geometric quantization of the group $SU(2)$. We now comment briefly on the connections between the Bargmann wavefunctions and the holomorphic wavefunctions constructed from geometric quantization of $SU(2)$. The wavefunctions that result from the geometric quantization of the coadjoint orbits of $SU(2)$, which are $2$-spheres, are holomorphic functions on those coadjoint orbits. These holomorphic functions can be obtained from the Bargmann wavefunctions by restricting them to a  ${\mathbb C}P^1 = S^2$ subspace of ${\mathbb C}^2$.  

Let us focus on a single angular momentum. We start with the Bargmann wavefunction for the basis state $\ket{j m}$, given by 

\begin{equation}
\psi_{jm} (z_1, z_2) = \frac{1}{\sqrt{(j-m)! (j+m)!}} \, z^{j+m} \, z^{j-m} \, .
\end{equation}
The function $\psi_{jm}(z_1, z_2)$ has the scaling property 

\begin{equation}
\psi_{jm} (\lambda z_1, \lambda z_2) = \lambda^{2j} \, \psi_{jm} (z_1, z_2)  \, ,
\end{equation}
for  $\lambda \in {\mathbb C}$. Thus the Bargmann wavefunctions are completely determined by its values on a  ${\mathbb C}P^1 = S^2$ subspace. Let us choose this subspace to be  defined by $|z_1|^2 + |z_2|^2 = 1$, and set $z_1$ real. Let $\xi = z_2 / z_1$ parametrize the coordinate on this subspace. Then the basis functions $\psi_{jm} (z_1, z_2)$ restricted to the section ${\mathbb C}P^1$ is given by 

\begin{eqnarray}
\psi_{jm} (\xi) &=& \frac{ z_1^{j+m} \, z_2^{j-m} }{[ (j+m)! \, (j-m)! \,]^{1/2}}  \nonumber  \\  \nonumber
&=& \frac{z_1^{2j} ( z_2 / z_1 )^{j-m} }{[ (j+m)! \, (j-m)! \,]^{1/2}}   \\
&=& \frac{1}{[ (j+m)! \, (j-m)! \,]^{1/2}} \frac{\xi^{j-m}}{(1 + |\xi|^2)^j} \, , 
\end{eqnarray}
where we have used $ 1 = |z_1|^2 + |z_2|^2 = z_1^2 ( 1 + |\xi|^2 )$. These basis wavefunctions $\psi_{jm}(\xi)$ agree with  those used in geometric quantization of $SU(2)$  on page 177 of Woodhouse\cite{woodhouse1991}.

\section{\label{ch2: sec_rel_schwinger}Relation to the Schwinger Representation}

Another recent approach\cite{littlejohn2007, littlejohn2010a} to the semiclassical analysis of the Wigner $3nj$-symbols uses the WKB wavefunctions of the harmonic oscillators in Schwinger's model of angular momentum. It turns out the Bargmann wavefunctions are the exact harmonic oscillator states for those WKB wavefunctions in the coherent state representation. To show this, we write the eigenstates of the harmonic oscillators in the coherent states basis of the Heisenberg-Weyl group \cite{perelomov1986}. The coherent states basis are given by $\ket{z} = \hat{D}(z) \ket{0}$,  where $\ket{0}$ is the ground state of the simple harmonic oscillator, and $\hat{D}(z)$ is an element of the Heisenberg-Weyl group, given by 

\begin{equation}
\hat{D}(z) = \exp  \left[ \frac{i}{\hbar} ( x \hat{p} - p \hat{x} ) \right] = \exp (- |z|^2 /2 ) \exp(z \hat{a}^\dagger) \exp( - \overline{z} \hat{a} ) \, , 
\end{equation}
where $z = x + i p$, and $\hat{a}$ and $\hat{a}^\dagger$ are the usual annihilation and creation operators of the simple harmonic oscillator. The coherent state wavefunction

\begin{equation}
\braket{x | z } = ( \pi \hbar)^{-1/4} \, \exp \left[ - \frac{z^2}{2} + (2 / \hbar)^{1/2} z x - \frac{x^2}{2 \hbar}  \right] 
\end{equation}
is the displaced ground state of the simple harmonic oscillator centered at $(x, p)$. Using the exponential generating function for the Hermite polynomials 

\begin{equation}
\exp(-z^2 + 2 z x )  = \sum_{n=0}^\infty \, \frac{z^n}{n!} \, H_n (x) \, , 
\end{equation}
we find that the eigenstates $\ket{n}$ of the simple harmonic oscillator written in the coherent state basis is given by 

\begin{equation}
\braket{z | n } = \sum_{n'}   \frac{z^{n'}}{n'!}  \int dx  ( \pi \hbar)^{-1/4} H_{n'}(x) e^{-x^2 / 2 \hbar} \psi_n (x) = \frac{z^n}{n!} \, , 
\end{equation}
where we have used the orthonormality of the eigenstates of the simple harmonic oscillator.

Schwinger's representation uses two simple harmonic oscillators. The basis states $\ket{j_1 m} = \ket{n_1 n_2}$ is the product of two eigenstates of the two harmonic oscillators, where $n_1 = j + m$ and $n_2 = j- m$. Thus, the Schwinger wavefunction for $\ket{j m}$ written in the coherent state basis is given by

\begin{equation}
\braket{z_1 z_2 | n_1 n_2 } = \frac{z_1^{j+m} z_2^{j-m}}{(j+m)! (j-m)!} \, , 
\end{equation}
which is proportional to the Bargmann wavefunction $\psi_{jm}(z) = \braket{z| j m}$. Thus, the Bargmann wavefunctions are the exact states of the WKB wavefunctions from the multidimensional approach.

\section{Conclusions}

In this paper, we have provided a new derivation for the asymptotic formula of the Wigner $3j$-symbol, using exact wavefunctions in the Bargmann representation. In addition, we have pointed out the relationship between the Bargmann wavefunctions and other asymptotic wavefunctions used in two other recent approaches to the semiclassical analysis of the Wigner $3nj$-symbols.

In the derivation in the this paper, as well as in the other two approaches, the stationary phase points have natural geometrical interpretation in terms of classical vectors. The Bargmann approach and the geometric quantization approach present easy ways to calculate the asymptotic phase, whereas the multidimensional WKB approach provides an elegant formula for the amplitude determinants. The Bargmann approach, however, is the only one that uses the exact wavefunctions directly. This may make it possible to calculate higher order terms in the asymptotic series.


\appendix

\section{\label{ch2: sec_stationary_spinors}Finding A Stationary Phase Point}

\begin{figure}[tbhp]
\begin{center}
\includegraphics[width=0.40 \textwidth]{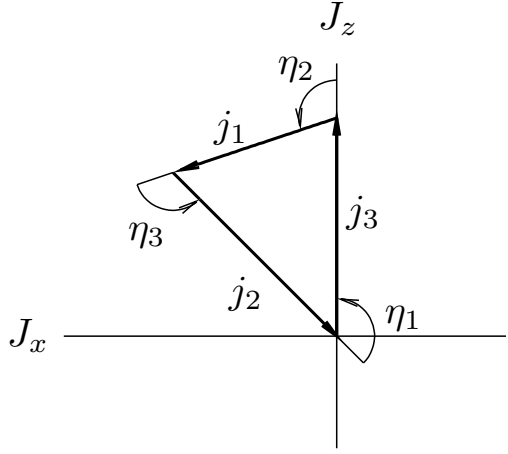}
\caption{Visualization of the stationary phase conditions $|{\mathbf J_r}| = j_r$, $r = 1,2,3$, and ${\mathbf J}_i = {\mathbf 0}$, $i = x, y, z$, as a triangle with edge lengths $j_1, j_2, j_3$.}
\label{ch2: fig_triangle}
\end{center}
\end{figure}

We now find one stationary phase point that satisfy Eq. (\ref{ch2: eq_stationary_phase_condition}). Let us start with a standard orientation for a triangle formed by ${\bf J}_1, {\bf J}_2, {\bf J}_3$ by placing ${\bf J}_3$ along the $z$-axis, as illustrated in Fig.\ \ref{ch2: fig_triangle}. The angles $\eta_r$ lie in the range $0 \le \eta_r \le \pi$. and are complementary to the interior angles of the triangle with lengths $j_1, j_2, j_3$. From the law of cosine for triangles, we find

\begin{equation}
\label{ch2: eq_eta}
\cos \eta_1 = \frac{j_1^2 - j_2^2 - j_3^2}{2j_2 j_3}  \, , \quad \quad 
\cos \eta_2 = \frac{j_2^2 - j_1^2 - j_3^2}{2j_1 j_3} \, ,
\end{equation}
and cyclic permutations.

In that orientation, the vectors in Euclidean coordinates are given by

\begin{equation}
\label{ch2: eq_ref_vectors}
{\bf J}_1 = j_1 
\left(
\begin{array}{c}
\sin \eta_2 \\
0 \\
\cos \eta_2
\end{array} \right) \, ,
\quad \quad 
{\bf J}_2 = j_2 
\left(
\begin{array}{c}
- \sin \eta_1 \\
0 \\
\cos \eta_1
\end{array} \right) \, ,
\quad \quad 
{\bf J}_3 = j_3 
\left(
\begin{array}{c}
0 \\
0 \\
1
\end{array} \right) \, .
\end{equation}

We find the spinors on the Hopf fiber above these vectors. Because all three vectors lie in the $xz$ plane, we can choose all three spinors to be real. We choose

\begin{eqnarray}
\label{ch2: eq_ref_spinor}
&& \left( 
\begin{array}{c}
z_{11}  \\
z_{12} 
\end{array}  \right)
= \sqrt{2j_1}
\left(
\begin{array}{c}
\cos \eta_2 /2  \\
\sin \eta_2 /2
\end{array}  \right),
\quad \quad 
\left( 
\begin{array}{c}
z_{21}  \\
z_{22} 
\end{array}  \right)
= \sqrt{2j_2}
\left(
\begin{array}{c}
\cos \eta_1 /2  \\
- \sin \eta_1 /2
\end{array}  \right) ,   \\   \nonumber
&&  \left( 
\begin{array}{c}
z_{31}  \\
z_{32} 
\end{array}  \right)
= \sqrt{2j_3}
\left(
\begin{array}{c}
1  \\
0
\end{array}  \right) .
\end{eqnarray}

\begin{figure}[tbhp]
\begin{center}
\includegraphics[width=0.40 \textwidth]{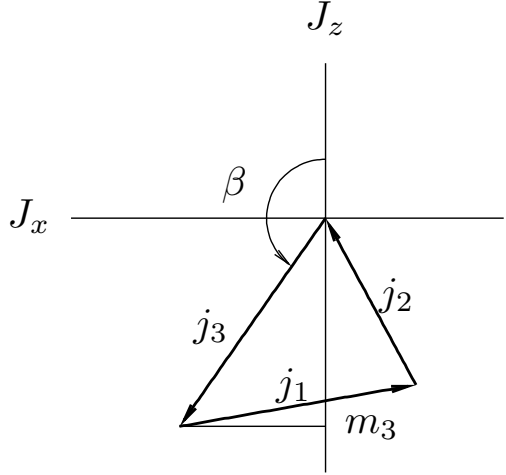}
\caption{The vectors after applying the rotation about the $y$-axis by an angle $\beta$ to the triangle in the reference orientation.}
\label{ch2: fig_beta}
\end{center}
\end{figure}

We now apply rotations to the vectors in Eq.\ (\ref{ch2: eq_ref_vectors}) to ensure their $z$ projections satisfy $J_{rz} = m_r$, $r = 1,2,3$. We do this in two steps. First we rotate the vectors in the $x$-$z$ plane about the $y$-axis by an angle $\beta$, $0 \le \beta \le \pi$, defined by 

\begin{equation}
\label{ch2: eq_beta}
m_3 = j_3 \cos \beta  \, , 
\end{equation}
so that $J_{3z} = m_3$. This rotation is illustrated in Fig.\ \ref{ch2: fig_beta}.

Once we have $J_{3z} = m_3$, we rotate the vectors about the vector ${\mathbf J}_3$ by an angle $\gamma$ to satisfy $J_{2z} = m_2$. This rotation preserves $J_{3z} = m_3$. Moreover, because $J_{1z} + J_{2z} + J_{3z} = 0$, we get $J_{2z} = - m_1 - m_3 = m_2$ automatically.

To find the angle $\gamma$, we calculate the final vectors of applying the rotations $R({\mathbf j}_3, \gamma) \, R({\mathbf y}, \beta) =  R({\mathbf y}, \beta) \, R({\mathbf j}_3, \gamma)$ to the vector in Eq.\ (\ref{ch2: eq_ref_vectors}),

\begin{eqnarray}
{\bf J}_1 &=& J_1 
\left(
\begin{array}{c}
\cos \beta \cos \gamma \sin \eta_2 + \sin \beta \cos \eta_2  \\
\sin \gamma \sin \eta_2 \\
- \sin \beta \cos \gamma \sin \eta_2 + \cos \beta \cos \eta_2
\end{array} \right) \, ,   \nonumber  \\
\label{ch2: eq_3j_vec_sol}
{\bf J}_2 &=& J_2
\left(
\begin{array}{c}
- \cos \beta \cos \gamma \sin \eta_1 + \sin \beta \cos \eta_1 \\
- \sin \gamma \sin \eta_1 \\
\sin \beta \cos \gamma \sin \eta_1 + \cos \beta \cos \eta_1
\end{array} \right) \, ,    
\end{eqnarray}

\begin{eqnarray}
{\bf J}_3 &=& J_3
\left(
\begin{array}{c}
\sin \beta \\
0 \\
\cos \beta
\end{array} \right) \, .  \nonumber
\end{eqnarray}
Then impose the condition $J_{1z} = m_1$ on the final vectors to determine $\gamma$, 

\begin{equation}
\label{ch2: eq_gamma}
\cos \gamma = \frac{j_1 \cos \beta \cos \eta_2 - m_1}{j_1 \sin \beta \sin \eta_2}  \, . 
\end{equation}
In general, we find two solutions to Eq.\ (\ref{ch2: eq_gamma}). Let $\gamma$ represent the root of (\ref{ch2: eq_gamma}) in the range $[0,  \pi]$, and $- \gamma$ the root  in the range $[ - \pi , 0]$. 

We now lift the rotations $R({\mathbf y}, \beta) \, R({\mathbf j}_3, \gamma)$ up to an $SU(2)$ rotation 

\begin{equation}
\label{ch2: eq_spinor_rotation}
u({\bf y}, \beta) \, u({\bf z}, \gamma) = 
\left(
\begin{array}{cc}
e^{-i \gamma / 2} \cos \beta / 2  &  - e^{i \gamma /2} \sin \beta / 2  \\
e^{-i \gamma / 2} \sin \beta / 2  &  e^{i \gamma / 2}  \cos \beta / 2
\end{array}
\right) \, , 
\end{equation}
and apply this spinor rotation (\ref{ch2: eq_spinor_rotation}) to the reference spinors (\ref{ch2: eq_ref_spinor}). This gives a set of spinors that project onto the final vectors in Eq.\ (\ref{ch2: eq_3j_vec_sol}). This way, we have found one point $p$ that satisfy the stationary phase conditions. The result is displayed in Eq.\ (\ref{ch2: eq_final_spinor_1}).

\section{\label{ch2: sec_hessian}The Calculation of the Hessian}

The stationary phase points consist of two $4$-tori, so we pick eight transversal directions to calculate the Hessian. Let $z_{s \mu} = r_{s\mu} e^{i \phi_{s\mu}}$, $s = 1,2,3$, $\mu=1,2$,  be the complex coordinates written in polar coordinates.  Introduce the new variables

\begin{equation}
Z_s = z_{s1} / z_{s2} = R_s e^{i \Phi_s}  \, ,
\end{equation}
where $R_s = r_{s1} / r_{s2}$ and $\Phi_s = \phi_{s1} - \phi_{s2}$. 

We choose the eight directions transversal to the stationary phase points to be $\partial_{\Phi_1}$, $\partial_{\Phi_2}$, $\partial_{R_1}$, $\partial_{R_2}$, $\partial_{R_3}$, $\partial_{r_{12}}$, $\partial_{r_{22}}$, $\partial_{r_{32}}$. This is not an orthonormal coordinate system, so the change of coordinates generates a Jacobian factor $\prod_{s\mu} r_{s\mu}^{-2}$ in the Hessian.

Using the new variables $Z_s$, the phase function $f$ from Eq.\ (\ref{ch2: eq_phase_function}) can be rewritten as 

\begin{eqnarray}
f &=&  \sum_s \, \left[ (j_s + m_s) \, \ln \overline{Z}_s \right]
   + k_1 \, \ln ( Z_2 - Z_3 ) + k_2 \, \ln ( Z_3 - Z_1 )   \\   \nonumber
 && \quad \quad   + k_3 \, \ln ( Z_1 - Z_2 ) 
  - \sum_s r_{s2}^2 R_s^2  
 +  \sum_s \, 4 j_s \ln r_{s2}  - \sum_s \, r_{s2}^2  \, . 
\end{eqnarray}
Its first derivatives are displayed in Eqs.\ (\ref{ch2: eq_condition_1}) - (\ref{ch2: eq_f_derivative_last}). Out of the $36$ distinct second derivatives of $f$, there are $21$ nonzero ones. These are listed in in Eqs.\ (\ref{ch2_second_derivative_begin}) - (\ref{ch2_second_derivative_end}).

\begin{eqnarray}
\frac{\partial \, f}{\partial \Phi_1} 
	&=& i \left[ - (j_1 + m_1) + \frac{ -  k_2 Z_1}{Z_3 - Z_1} + \frac{ k_3 Z_1}{Z_1 - Z_2}  \right]  \, ,  \label{ch2: eq_condition_1}   \\
\frac{\partial \, f}{\partial \Phi_1} 
	&=&  i \left[ - (j_2 + m_2) + \frac{ k_1 Z_2}{Z_2 - Z_3} + \frac{-  k_3 Z_2}{Z_1 - Z_2}  \right]   \, ,   \label{ch2: eq_condition_2}    \\
\frac{\partial \, f}{\partial R_1} 
	&=& \frac{1}{R_1} \left[ (j_1 + m_1) + \frac{- k_2 Z_1}{Z_3 - Z_1} + \frac{k_3 Z_1}{Z_1 - Z_2} -  2 r_{12}^2 \, R_1^2
\right]  \, ,   \\
\frac{\partial \, f}{\partial R_2} 
	&=& \frac{1}{R_2} \left[ (j_2 + m_2) + \frac{ k_1 Z_2}{Z_2 - Z_3} + \frac{- k_3 Z_2}{Z_1 - Z_2} -  2 r_{22}^2 \, R_2^2
\right]  \, ,  \\
\frac{\partial \, f}{\partial R_3} 
	&=& \frac{1}{R_3} \left[ (j_3 + m_3) + \frac{- k_1 Z_3}{Z_2 - Z_3} + \frac{k_2 Z_3}{Z_3 - Z_1} -  2 r_{32}^2 \, R_3^2
\right]  \, ,   \\
\frac{\partial \, f}{\partial r_{12}} 
	&=& - 2 ( 1 + R_1^2 ) r_{12} + \frac{4j_1}{r_{12}}  \, ,  \\
\frac{\partial \, f}{\partial r_{22}} 
	&=& - 2 ( 1 + R_2^2 ) r_{22} + \frac{4j_2}{r_{22}}  \, ,   \\
\frac{\partial \, f}{\partial r_{32}} 
	&=& - 2 ( 1 + R_3^2 ) r_{32} + \frac{4j_3}{r_{32}}  \, . 
	\label{ch2: eq_f_derivative_last}
\end{eqnarray}

\begin{eqnarray}
\label{ch2_second_derivative_begin}
\frac{\partial^2  \, f}{\partial \Phi_1 \, \partial \Phi_1}  
    	&=&   \frac{ k_2 Z_1 Z_3 }{(Z_3 - Z_1)^2} + \frac{ k_3 Z_1 Z_2 }{(Z_1 - Z_2)^2}     \\   
\frac{\partial^2  \, f}{\partial \Phi_1 \, \partial \Phi_2}  
    	&=&  - \frac{k_3 Z_1 Z_2}{(Z_1 - Z_2)^2}  \\    	
 \frac{\partial^2  \, f}{\partial \Phi_1 \, \partial R_1}  
    	&=&   - i  \; \left[  \frac{k_2 Z_3 e^{i \Phi_1} }{(Z_3 - Z_1)^2} + \frac{ k_3 Z_2 e^{i \Phi_1}}{(Z_1 - Z_2)^2}  \right]   \\   
\frac{\partial^2  \, f}{\partial \Phi_1 \, \partial R_2} 
    	&=&  \frac{ i k_3 Z_1 e^{i \Phi_2} }{( Z_1 - Z_2 )^2 }     \\
 \frac{\partial^2  \, f}{\partial \Phi_1 \, \partial R_3}  
    	&=&  \frac{ i k_2 Z_1 }{( Z_3 - Z_1 )^2 }    \\    
 \frac{\partial^2  \, f}{\partial \Phi_2 \, \partial \Phi_2} 
    	&=&  \frac{ k_1 Z_2 Z_3 }{(Z_2 - Z_3)^2} + \frac{ k_3 Z_1 Z_2 }{(Z_1 - Z_2)^2}    \\
 \frac{\partial^2  \, f}{\partial \Phi_2 \, \partial R_1}  
	&=&    \frac{ i k_3 Z_2 e^{i \Phi_1} }{( Z_1 - Z_2 )^2 }   
\end{eqnarray}

\begin{eqnarray}	
 \frac{\partial^2  \, f}{\partial \Phi_2 \, \partial R_2} 
	&=&  -i \left[  \frac{k_1 \, Z_3 \, e^{i \Phi_2} }{(Z_2 - Z_3)^2} + \frac{ k_3 \,  Z_1 \, e^{i \Phi_2}}{(Z_1 - Z_2)^2}  \right]     \\	
 \frac{\partial^2  \, f}{\partial \Phi_2 \, \partial R_3}   
	&=&   \frac{ i k_1 Z_2 }{( Z_2 - Z_3 )^2 }     \\   
 \frac{\partial^2 f}{ \partial R_1 \partial R_1 }  
	&=&  -  \left[ \frac{k_2 \, Z_3 \, e^{i \Phi_1}}{R_1 \; ( Z_3 - Z_1 )^2 }  + \frac{k_3 \, Z_2 \, e^{i \Phi_1}}{R_1 \; ( Z_1 - Z_2 )^2 } +  4 \,  r_{12}^2  \right]       \\  
 \frac{\partial^2 f}{ \partial R_1 \partial R_2 }   
	&=&   \frac{ k_3 \; e^{i \Phi_1} \; e^{i \Phi_2} }{ (Z_1 - Z_2)^2 }     \\    
 \frac{\partial^2 f}{ \partial R_1 \partial R_3 }  
	&=&   \frac{ k_2 \; e^{i \Phi_1}}{(Z_3 - Z_1)^2}    \\    
 \frac{\partial^2 f}{ \partial R_2 \partial R_2 }  
	&=&   -  \left[ \frac{k_1 \, Z_3 \, e^{i \Phi_2}}{R_2 \; ( Z_2 - Z_3 )^2 } + \frac{k_3 \, Z_1 \, e^{i \Phi_2}}{R_2 \; ( Z_1 - Z_2 )^2 } +  4 \,  r_{22}^2  \right]     \\    
 \frac{\partial^2 f}{ \partial R_2 \partial R_3 } 
	&=&  \frac{ k_1 \; e^{i \Phi_2} }{ (Z_2 - Z_3)^2 }    \\   
  \frac{\partial^2 f}{ \partial R_3 \partial R_3 }    
	&=&  -  \left[ \frac{k_1 \, Z_2 }{R_3 \; ( Z_2 - Z_3 )^2 } + \frac{k_2 \, Z_1 }{R_3 \; ( Z_3 - Z_1 )^2 }  +  4 \,  r_{32}^2  \right]      \\
 \frac{\partial \, f}{\partial R_1 \, \partial r_{12}}   
	&=&  - 4 r_{12} R_1  	  \\   
 \frac{\partial \, f}{\partial R_2 \, \partial r_{22}}  
	&=&  - 4 r_{22} R_2     \\     
 \frac{\partial \, f}{\partial R_3 \, \partial r_{32}}    
	&=&  - 4 r_{32} R_3      \\      
 \frac{\partial \, f}{\partial r_{12}\, \partial r_{12}}  
	&=&   - 2 ( 1 + R_1^2 + \frac{2j_1}{r_{12}^2} )     \\     
 \frac{\partial \, f}{\partial r_{22} \, \partial r_{22}}   
	&=&   - 2 ( 1 + R_2^2   + \frac{2j_2}{r_{22}^2} )    \\     
\label{ch2_second_derivative_end}
 \frac{\partial \, f}{\partial r_{32} \, \partial r_{32}}     
	&=&   - 2 ( 1 + R_3^2 + \frac{2j_3}{r_{32}^2} )     
\end{eqnarray}

After simplifying the Hessian matrix through Gaussian eliminations on the last six rows and columns, which preserves the determinant, we find that the Hessian matrix has the form

\begin{equation}
H = 
\left( 
\begin{array}{cccccccc}
 H_{11}  & H_{12} & 0 & 0 & 0 & 0 & 0  \\
H_{21} & H_{22} & 0 & 0 &  0 & 0 &  0 &  0 \\
0 & 0 & H_{33} & 0 & 0 & 0 & 0 & 0 \\
0 & 0 & 0 &  H_{44} & 0 & 0 & 0 & 0 \\
0 & 0 & 0 & 0 &  H_{55} & 0 & 0 & 0 \\
0 & 0 & 0 & 0 & 0 & H_{66} & 0 & 0 \\
0 & 0 & 0 & 0 & 0 & 0 & H_{77} & 0 \\
0 & 0 & 0 & 0 & 0 & 0 & 0 & H_{88} \\
\end{array}
\right)  \, , 
\end{equation}
where

\begin{eqnarray}
H_{11} & = & \frac{\partial^2  \, f}{\partial \Phi_1 \, \partial \Phi_1}  \, ,  \quad \quad
H_{12}  =   H_{21} = \frac{\partial^2  \, f}{\partial \Phi_1 \, \partial \Phi_2}  \, ,  \quad \quad 
H_{22}  =  \frac{\partial^2  \, f}{\partial \Phi_2 \, \partial \Phi_2}   \, ,  \\
H_{33} & = &  -4 r_{12}^2  \, ,  \quad \quad  \quad \quad
H_{44}  =   - 4 r_{22}^2    \, ,   \quad \quad  \quad \quad
H_{55}  =  - 4 r_{32}^2 \, ,  \\
H_{66} & = &  - 2 ( 1 - R_1^2 + \frac{2j_1}{r_{12}^2} )  \, ,   \quad \quad  \quad \quad \quad 
H_{77}  =  - 2 ( 1 - R_2^2   + \frac{2j_2}{r_{22}^2} )  \, ,   \\
H_{88}  &=&  - 2 ( 1 - R_3^2 + \frac{2j_3}{r_{32}^2} )  \, .
\end{eqnarray}
At the stationary phase points, $r_{s1} = \sqrt{j_s+m_s}$, $r_{s2} = \sqrt{j_s - m_s}$, $s = 1,2,3$, so the last three diagonal entries are constants. 

\begin{eqnarray}
 H_{55} = H_{66} = H_{77}  
&=& - 2 ( 1 - R_s^2 + 2 j_s / r^2_{s2} ) \\  \nonumber 
&=& - 2 \left( 1 - \frac{j_s + m_s}{j_s - m_s} + \frac{2 j_s}{j_s - m_s} \right)   \\  \nonumber 
&=& -4 \,  . 
\end{eqnarray}
Taking the determinant, we find

\begin{eqnarray}
&&\det H =  4^6  \left[ \frac{k_1 \, k_2 \, Z_1 \, Z_2 \, Z_3^2}{(Z_3 - Z_1)^2 \, (Z_2 - Z_3)^2} + 
         \frac{k_2 \, k_3 \, Z_1^2 \, Z_2 \, Z_3}{(Z_3 - Z_1)^2 \, (Z_1 - Z_2)^2}    \right.  \\   \nonumber 
&&   \quad \quad \quad \quad \quad   \left.	 +  \frac{k_1 \, k_3 \, Z_1 \, Z_2^2 \, Z_3}{(Z_1 - Z_2)^2 \, (Z_2 - Z_3)^2} \right]  \, . 
\end{eqnarray}
This determinant can be put into a more symmetrical form, 

\begin{eqnarray}
\det H &=&  4^6  \frac{Z_1 \, Z_2 \, Z_3 }{ (Z_1-Z_2) (Z_2-Z_3) (Z_3-Z_1) }  
  \left[ \frac{k_1 \, k_2 \, Z_3 \, (Z_1 - Z_2)}{(Z_3 - Z_1) \, (Z_2 - Z_3)}   \right.  \\   \nonumber   
&&  \quad  \quad \quad \quad 	\left.	            +  \frac{k_2 \, k_3 \, Z_1 \, (Z_2 - Z_3)}{(Z_3 - Z_1) \, (Z_1 - Z_2)}    
   +  \frac{k_1 \, k_3 \, Z_2 \, (Z_3 - Z_1)}{(Z_1 - Z_2) \, (Z_2 - Z_3)} \right]   \\   \nonumber   
&=& g \left[  \left( \frac{- k_2 Z_2 }{ Z_3 - Z_1 } + \frac{ k_3 Z_2 }{ Z_1 - Z_2 }   \right) 
								\left( \frac{k_1 Z_3 }{ Z_2 - Z_3 } + \frac{- k_3 Z_1 }{ Z_1 - Z_2 }   \right)  \right.  \\   \nonumber
&& \left.  - \left( \frac{k_2 Z_3 }{ Z_3 - Z_1 } + \frac{ - k_3 Z_2 }{ Z_1 - Z_2 }   \right)  							
	\left( \frac{- k_1 Z_1 }{ Z_2 - Z_3 } + \frac{k_3 Z_1 }{ Z_1 - Z_2 }   \right)   \right]  \\   \nonumber
&=& g [ g_1 g_2 - g_3 g_4 ]  \, , 
\end{eqnarray}
where $g$ corresponds to the factor in front, and $g_1, g_2, g_3, g_4$ correspond to the four factors in the four parenthesis, respectively. We now express them in terms of the variables $z_{r\mu}$. We find

\begin{eqnarray}
g &=& 4^6  \frac{Z_1 \, Z_2 \, Z_3 }{ (Z_1-Z_2) (Z_2-Z_3) (Z_3-Z_1) }  \\
&=& 4^6  \frac{z_{11} z_{12} z_{21} z_{22} z_{31} z_{32} }{(z_{11} z_{22} - z_{12}z_{21}) (z_{31} z_{12} - z_{32}z_{11}) (z_{21} z_{32} - z_{22}z_{31}) }  \, . 
\end{eqnarray}
Setting the first derivatives (\ref{ch2: eq_condition_1}) and (\ref{ch2: eq_condition_2}) to zero, and using $|z_{r1}|^2 = j_r + m_r$,  we can rewrite $g_1$ and $g_4$ as

\begin{equation}
g_1 = \frac{- k_2 Z_2 }{ Z_3 - Z_1 } + \frac{ k_3 Z_2 }{ Z_1 - Z_2 } =
	( j_1 + m_1 ) \frac{Z_2}{Z_1} 
	= z_{11} \overline{z}_{11}  \frac{z_{21} z_{12}}{z_{22} z_{11}} 
	= \frac{z_{21} z_{12} \overline{z}_{11}}{z_{22}}   \, , 
\end{equation}
\begin{equation}
g_4 = \frac{- k_1 Z_1 }{ Z_2 - Z_3 } + \frac{k_3 Z_1 }{ Z_1 - Z_2 }  	=
	( j_2 + m_2 ) \frac{Z_1}{Z_2} 
	= z_{21} \overline{z}_{21}  \frac{z_{11} z_{22}}{z_{12} z_{21}} 
	= \frac{z_{11} z_{22} \overline{z}_{21}}{z_{12}}  \, .  \\
\end{equation}
Similarly, setting the first derivatives (\ref{ch2: eq_condition_1}) and (\ref{ch2: eq_condition_2}) to zero, and using $2j_1 = k_2 + k_3$, $2j_2 = k_1 + k_3$, and $|z_{r2}|^2 = j_r - m_r$, rewrite $g_2$ and $g_3$ as

\begin{eqnarray}
g_2 &=& \frac{k_1 Z_3 }{ Z_2 - Z_3 } + \frac{- k_3 Z_1 }{ Z_1 - Z_2 }   \nonumber 
	=  \frac{k_1 Z_2 }{ Z_2 - Z_3 } + \frac{- k_3 Z_2 }{ Z_1 - Z_2 } - (k_1 + k_3)  \\  \nonumber
&=&  (j_2 + m_2) - 2 j_2 
= -  (j_2 - m_2)  \\
&=&  -  z_{22} \overline{z}_{22} \, , 
\end{eqnarray}
\begin{eqnarray}
g_3 &=& \frac{k_2 Z_3 }{ Z_3 - Z_1 } + \frac{ - k_3 Z_2 }{ Z_1 - Z_2 }  \nonumber
	= \frac{k_2 Z_1 }{ Z_3 - Z_1 } + \frac{ - k_3 Z_1 }{ Z_1 - Z_2 } + ( k_2 + k_3 )  \\  \nonumber
&=& - (j_1 + m_1) + 2 j_1
= (j_1 - m_1)  \\   \nonumber
&=& z_{12} \overline{z}_{12}  \, .
\end{eqnarray}
Using the values of $g_1, g_2, g_3, g_4$, and $g$ above, we find

\begin{eqnarray}
\det H &=& g ( g_1 g_2 - g_3 g_4 )    
= g ( -  z_{21} z_{12} \overline{z}_{11} \overline{z}_{22} + z_{11} z_{22} \overline{z}_{12} \overline{z}_{21} )   \nonumber   \\   \nonumber
&=& g [ - 2 i (J_{1x} J_{2y} - J_{2x} J_{1y} ]   \\
&=& - 4 i g \,  \Delta_{z}  \, , 
\end{eqnarray}
where we have used Eq.\ (\ref{ch2: eq_Hopf_Jrx}) - (\ref{ch2: eq_Hopf_Jrz}) in the third equality. In the last equality, we have defined

\begin{equation}
\label{ch2: eq_delta_z}
\Delta_z = \frac{1}{2} ( J_{1x} J_{2y} - J_{2x} J_{1y} )  \, . 
\end{equation}
It is the area of the triangle in the angular momentum space, projected onto the $x$-$y$ plane. Finally, the Hessian is equal to $\det H$ times the Jacobian $\prod_{s\mu} r_{s\mu}^{-2}$. The result is 

\begin{eqnarray}
{\rm Hessian}   
&=& (-i ) 4^7 \Delta_z \left[  \,  \frac{ \left( \prod_{s\mu} r_{s\mu}^{-2} \right)   \, z_{11} z_{12} z_{21} z_{22} z_{31} z_{32} }{(z_{11} z_{22} - z_{12}z_{21}) (z_{31} z_{12} - z_{32}z_{11}) (z_{21} z_{32} - z_{22}z_{31}) } \right]  \nonumber  \\   
&=&   \frac{(-i ) 4^7  \, \Delta_z}{\overline{z}_{11} \overline{z}_{12} \overline{z}_{21} \overline{z}_{22} \overline{z}_{31} \overline{z}_{32} (z_{11} z_{22} - z_{12}z_{21}) (z_{31} z_{12} - z_{32}z_{11}) (z_{21} z_{32} - z_{22}z_{31}) }  \, .
\label{ch2: eq_Hessian}
\end{eqnarray}
Note that the factor $-i$ is responsible for the Maslov index $1/\sqrt{-i} = e^{i \pi / 4}$.

\end{document}